# Alternating current driven interfacial dynamics of a binary fluid in a patterned fluidic environment modulated by electrothermal effects

Golak Kunti, Anandaroop Bhattacharya, Suman Chakraborty,[a]

[1]Department of Mechanical Engineering, Indian Institute of Technology Kharagpur, Kharagpur, West Bengal - 721302, India

[a]*E-mail address* of the corresponding author: suman@mech.iitkgp.ernet.in

**ABSTRACT**

In this paper, we depict interfacial electro-thermo-chemical-hydrodynamics of two immiscible fluids in a microchannel with substrates patterned by ribs. The motion of the binary fluids is set by an alternating current electrothermal (ACET) mechanism. Our investigations, based on free-energy-based phase field formalism, reveal that the capillary filling dynamics and the contact line motion are strong functions of the wetting characteristics and geometric parameters of the patterned ribs. Modulation of these parameters alters the surface energy over the rib surface, which, in turn, facilitates the interaction between interfacial tension and driving electrothermal forces. The competition between two forces may speed up or slow down the fluid-fluid-solid contact line motion over the rib surface. At the edges of the ribs, the interface can halt for a sufficiently long time owing to contact line pinning. Alteration in positions of the rib between the electrode pairs changes the electric field strength and thereby bulk ACET forces across the contact line. Further, by suitable arrangement of the ribs, various intricate shapes of the liquid front can be achieved over a short distance, which can have significant implications in the morphological control of microscale flows.

**I. Introduction**

Manipulation of biochemical reactions, biochemical analysis, particle deposition, clinical diagnostics, drug delivery, chemical synthesis are important processes related to the functionalities of microfluidic devices.[1–7] Two phase flow dynamics and its underlying physics, governing many of these applications, may therefore turn out to be of attractive interest to the research community.[8–10] When a fluid displaces another one along a solid surface, fluid-fluid-solid contact line is formed. The underlying morpho-dynamics may find potential applications in the area of textile manufacturing, chemical technologies, polymer processing, oil recovery, manufacturing of photographic films, powder wetting, microscale biochemical analysis and biomedical processes.[11–16] The effective control of transport of binary fluids in those applications is a big challenge due to integration and coupling of simultaneously occurring processes.

Most of the studies on contact line dynamics reported previously were based on pressure driven flow where mechanical actuators control the fluid flow.[17–19] Several issues of effectiveness and reliability, such as acoustic noise, moving part etc. limit the applicability of mechanical actuators in microfluidic devices. Recently, electric field driven flow actuation has drawn attention as an attractive driving mechanism. Chip scale integrability, operation without moving components and noise free environment enhance the efficiency and portability of electric field driven miniaturized systems.[20–22] Under electrical actuators, electrokinetic mechanisms have emerged broadly to control the microfluidic processes. Direct current (DC) as well as alternating current (AC) electrokinetics play an important role in micromixing, fluid pumping, particle sorting etc.[23–27] However, DC electrokinetics causes some detrimental effects due to application of high voltage (up to kV). Electrolytic reaction, high Joule heat, electrode degradation are the adverse effects of DC sources which may alter the function and decrease the effectiveness of the microfluidic devices.[28–30] However, AC electrokinetics employs pair of electrodes with adjacent electrodes over a range of frequency which allows generation of a high non-uniform electric field at the expense of small applied voltage (1-20V)f Therefore, electrolysis and formation of bubbles are reduced. AC electroosmosis (ACEO) uses low electrical conductivity (up to 1 mS/m) of sample and lower frequency (up to 100 kHz) to generate mobile charges over the electrodes whose movement into the bulk solution invokes flow motion .[31,32] At relatively high operating frequency and conductivity, electric charges are not accumulated in the electrical double layer (EDL) and EDL becomes thinner which screens the electrode potential.[33] As a result, ACEO forces are almost negligible in those ranges of operating parameters. On the other hand, AC electrothermal process works over a wide range of electrical parameters and overcomes the demerits of ACEO mechanism. When AC voltages are applied at high frequency ($>10^5$ Hz [31,34]) and high electrical conductivity ($>0.084$ S/m [35]), the interaction between electrical field and thermal field caused by Joule heat results in local gradient in fluid electrical properties, such as permittivity and electrical conductivity. The inhomogeneities of the properties in the AC field generates volumetric forces, named as electrothermal forces which drive the fluid in a microfluidic channel.[31,32,36,37]

Investigations of capillary filling dynamics, controlling filling length, estimation of filling time, controlling of advancing liquid front (ALF) are important areas of research for fabrication and design of modern microfluidic devices.[38–41] On the other hand, various microfluidic systems are fabricated with integrated microstructures in order to perform various functions, like increasing surface area, preventing roof collapse, enhancing capillary force, altering hydrodynamic streamline, changing fluid meniscus, increasing mixing efficiency, to name a few.[42–45] These microstructures, such as ribs, grooves, ridges, posts alter the capillary filling dynamics in an intriguing manner.[46–48] Filling rate may decrease or increase depending on patterning of the microstructures which can also generate desired shape of the ALF.[49] In a diagnostic sensor based system, concaved-shaped ALF enhances the rate of interaction between the antigen and antibody and increases the coefficient of variation.[49] In addition, speed and shape of the ALF play an important role to maintain a sufficient delaying time, to stop the motion and for avoiding air-bubble trap.[50,51] Two important features of flow control i.e., delaying time and speed acceleration-deceleration are smoothly handled by incorporating microstructures inside the channel. Recently, on alteration of electrical parameters of electroosmotic actuation, it was found that interface velocity of two immiscible fluids can be increased or decreased inside a capillary whose surfaces were wetted.[52,53] The interfacial motion can be stopped at specific locations fixing suitable values of flow actuation parameters.

So far, electrothermal flow based on fluid pumping, mixing of samples, particle manipulation were investigated through microchannel without any undulation of the channel surface.[25–27,37,54,55] The influences of the heterogeneous surface on fluid motion between electrode patterning of ACET process become much significant. Especially when immiscible fluids are transported, interfacial dynamics get altered intriguingly. Here, we investigate the contact line dynamics of a binary system over a patterned fluidic substrate, driven by alternating current electrothermal flow. The rib surfaces and the walls are decorated with predefined wettability conditions. The wetting characteristics have significant effects on interfacial motion. With the alteration of chemical conditions and geometric parameters of the ribs, such as the height, width and position, desired velocity and delaying time of interface can be obtained. The locations of the ribs change the distribution of the potential lines and, hence, alter the net driving forces. Finally, we arrange the ribs asymmetrically over bottom and top surface to get the desired shape of the interface which has important implications towards achieving a designed morphological evolution in a confined fluidic environment.

## II. Simulation setup and numerical methodology

A. Physical system

The computational setup is shown in the Fig. 1 which consists of a channel having walls patterned with equispaced microribs. The channel runs along the x direction with length and height $L$ and $H$, respectively. The various dimensions of the rib, such as height ($h$), width ($w$) and position ($x_c$, left end to mid-width) are shown in the schematic. The channel height runs along the y direction and the channel width is sufficiently large to consider the validity of two-dimensional analysis. Initially, two immiscible fluids A and B are introduced at left and right sides of the solution domain. The channel walls and the rib surfaces are wetted with predefined wettability condition (manifested by static contact angle $\theta_s$). Eight ribs are paved on the top and bottom walls symmetrically in between 10 pairs of electrodes. The various dimensions of one pair of electrodes are shown in the dotted box. We adopt coplanar asymmetric array of electrodes with electrode width $d_1$, electrode gap $g \, (= d_1)$ and wider electrode width $d_2$ which cause unidirectional fluid flow (along x direction).[25,54,56,57] The gap between two electrode pairs is $2s$. Before imposing the electric field, the interface is held at the location $L_1$. When the AC signal is applied on the electrodes, electric field and Joule heat are generated which cause variation on permittivity and electrical conductivity, which, in turn, generate ACET forces into the bulk fluid. The interface starts moving over the wetted surface and ribs toward the end of the channel. We use subscripts "A" and "B" to denote the properties of the fluids A and B, respectively. In the subsequent subsections, we discuss the details of the phase field model and governing equations.

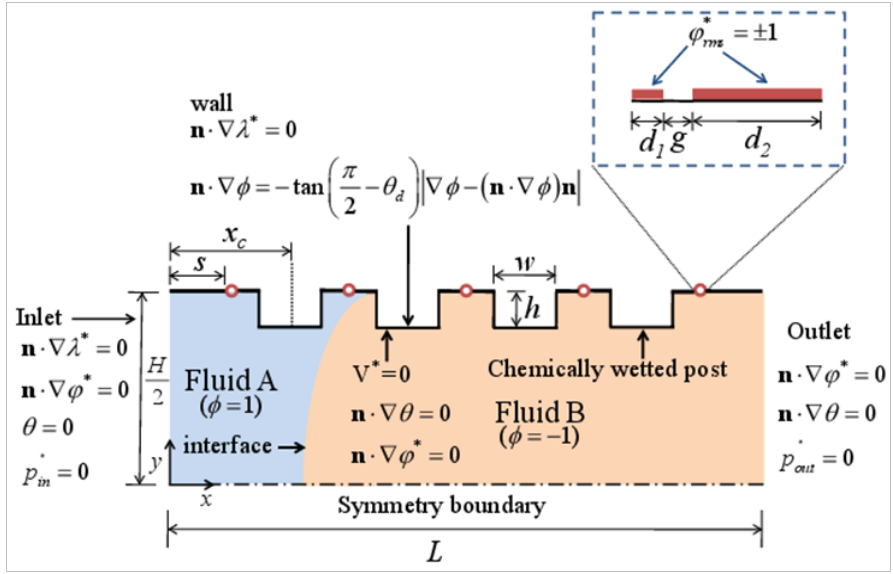

**Fig. 1.** Schematic representation of the physical system. The interface and the location of the fluids are shown. Ribs are patterned on the both the walls. Due to symmetry of the present model, we show only upper half portion. The origin is located at the middle of the left side. Various dimensions of the electrodes and ribs are shown. The boundary conditions are mentioned at appropriate places. On application of AC electric field on the electrodes, the interface starts moving along the x-direction, under the competition of electrothermal, viscous, and surface tension forces.

## B. Theory

Details description of governing transport equations and numerical methodology can be found in our previous study (Ref. [41]). For sake of completeness, we briefly highlight the numerical modeling in the following. The phase field model is the fundamental basis of thermodynamic systems where equilibrium states are identified on minimizing free energy of the multi phase system. [58–60] In our analysis, we use diffuse interface framework of phase field model to characterize the different states of the immiscible fluids. This method deals with phase field parameter $\phi$, commonly known as order parameter, defines the phase concentrations of the fluids. The order parameter is expressed in the form of number density ($n$) of the fluids, $\phi = (n_A - n_B)/(n_A + n_B)$, where $n_A$ and $n_B$ are for fluid A and fluid B, respectively. Thus, the fluids A and B can be denoted by the order parameter values $\phi = 1$ and $\phi = -1$, respectively, and the intermediate value $\phi \in (1, -1)$ indicates the interface which separates the two immiscible fluids. The Ginzburg-Landau free energy which describes the thermodynamics of the binary system can be written as. [61–66]

$$F = \int_\forall \left\{ f(\phi) + \frac{1}{2}\eta\xi|\nabla\phi|^2 \right\} d\forall, \qquad (1)$$

where $\forall$ is the entire volume of the system. $\eta$ and $\xi$ are the surface tension and interface thickness, respectively. In the integral, $f(\phi)$ is the free energy density in the bulk and the other term is the free energy density of the interface having a finite value. This term takes account the appropriate thickness of the interface separating the two phases while $f(\phi)$ maintains the immiscibility of the fluids and can be written in terms of double-well as [60,67] $f(\phi) = \frac{\eta}{4\xi}(1-\phi^2)^2$. The time evolution of the phase field variable is represented by the Cahn-Hilliard equation (CHE) which is the consequences of the mass conservation of the two phases and minimization of the free energy function as mentioned in the Eq.(1). The advection and diffusion of the order parameter is described by CHE as[60,67]

$$\frac{D\phi}{Dt^*} = \frac{1}{Pe_\phi}\nabla^2 \lambda^*, \qquad (2)$$

where $Pe_\phi = u_0 H^2 / M_c \eta$ is the phase field Peclet number. $H$ is the reference length. $M$ is the mobility parameter, also known as Onsager coefficient. Further, $M$ can be expressed in the form of critical mobility $M_c$ [62] and order parameter as $M = M_c(1-\gamma\phi^2)$. The parameter $\gamma$ controls the dynamical behavior of the interface of the binary system. The parameters with superscript * are taken as dimensionless form of the corresponding parameters. The chemical potential can be expressed as $\lambda^* = -Cn\nabla^2\phi + \frac{1}{Cn}(\phi^3 - \phi)$, where $Cn$ ($\xi/H$) is the Cahn number. We have taken properties of the fluid A as base fluid (denoted by subscript 0). The reference velocity is taken as $u_0 = \varepsilon_0 \varphi_0^2 \Delta T \beta / \mu_0 H$, $\Delta T$ is the maximum temperature difference in the system. Since maximum temperature occurs in the electrode gap, $\Delta T$ is the temperature difference between the temperature at electrode gap and reference temperature ($T_0$). $\varepsilon_0, \varphi_0$ and $\beta$ are the permittivity, voltage and gradient of the conductivity, respectively. For $\varepsilon_0 \approx 10^{-10}$ C/Vm, $\varphi_0 \approx 1$ V, $\Delta T \approx 1$ K, $\beta \approx 0.01$ K$^{-1}$, viscosity $\mu_0 \approx 0.001$ Pa s and $H \approx 10^{-4}$ m, we obtain $u_0 \approx 10^{-5}$ m/s. Various dimensionless parameters involved in our analysis are $t^* = \frac{tu_0}{H}, \theta = \frac{T-T_0}{\Delta T}, \varphi^* = \frac{\varphi}{\varphi_0}, p^* = \frac{pH}{\mu u_0}, \lambda^* = \frac{\lambda}{\eta/H}$, and $\omega^* = \frac{\omega}{\omega_0}$. Here, $t$ and $T$ are the time and temperature, respectively. $p$ and $\omega$ are the pressure and applied frequency, respectively. The reference frequency $\omega_0$ is inverse of the charge relaxation time ($\sigma_0/\omega_0$). Typical in an AC electrothermal field the solution electrical conductivity $\sigma_0 \sim 0.1$ S/m and applied frequency $\omega \geq 10^5$ Hz, which result operating dimensionless frequency ($\omega^*$) of the order of $10^{-3}$.

ACET forces are the outcome of inhomogeneities of electrical properties which arise when a temperature gradient prevails in the samples. In the present analysis, we consider internally evoked Joule heat to take account of the thermal field. Due to the application of non-uniform AC field Joule heat is generated. The potential distribution with negligible magnetic field effects [68,69] is obtained by solving the following equation: [70,71]

$$\nabla \cdot \left( \sigma^* \nabla \varphi^* \right) = 0 \tag{3}$$

As mentioned ACET forces are the coupled effect of the electric and thermal fields which result in internal heat generation. The thermal field is impacted through the interplay of convection, diffusion components with heat generation which can be expressed by energy equation as:

$$\mathrm{Pe}_T \rho^* C_p^* \frac{D\theta}{Dt^*} = k^* \nabla^2 \theta + \frac{1}{2} J \sigma^* \left| \mathbf{E}^* \right|^2, \tag{4}$$

where $\rho^*, C_p^*$ and $k^*$ are the fluid density, specific heat and thermal conductivity, respectively. $Pe_T = \rho_0 C_{p0} u_0 H / k_0$ and $J = \sigma_0 \phi^2 / k_0 \Delta T$ are the thermal Peclet number and Joule number, respectively. We denote Joule heating by $\sigma^* \left| \mathbf{E}^* \right|^2$.

Although we employ AC electrokinetic mechanism to cause the flow motion high diffusive flux is generated on alteration of order parameter between two fluids according to the diffuse interface model and, hence, surface tension force is generated. In addition, since we consider temperature gradient in the AC domain temperature dependent interfacial tension needs to be considered. The surface tension force and thermocapillary force combinedly can be written as [72] $\mathbf{F_S} = \frac{3\sqrt{2}}{4} \xi \left[ \eta_T \left| \nabla \phi \right|^2 \nabla T - \eta_T \left( \nabla T \cdot \nabla \phi \right) \nabla \phi + \frac{\eta}{\xi^2} \lambda \nabla \phi \right]$. $\eta_T$ is the gradient of the interfacial tension. A linear variation of surface tension with temperature is considered. Accordingly, surface tension can be expressed as $\eta(T) = \eta_0 + \eta_T (T - T_0)$. The interfacial tension force and electrothermal forces are added to the momentum equations. The modified Cahn-Hilliard-Navier-Stokes and continuity equations can be written as: [59,73,74]

$$\nabla \cdot \mathbf{V}^* = 0, \tag{5}$$

$$\mathrm{Re}\, \rho^* \frac{D\mathbf{V}^*}{Dt^*} = -\nabla p^* + \nabla \cdot \left[ \mu^* \left( \nabla \mathbf{V}^* + \nabla \mathbf{V}^{*T} \right) \right] + \frac{1}{\mathrm{Ca}} \mathbf{F_S}^* + \zeta \mathbf{F_E}^*, \tag{6}$$

Here, $\mathbf{V}^*$ is the fluid velocity vector. $\mathrm{Re} = \rho_0 u_0 H / \mu_0$ and $\mathrm{Ca} = \mu_0 u_0 / \eta$ are the Reynolds number and capillary number, respectively. $\zeta = \varepsilon_0 \phi^2 / \mu_0 u_0 H$ is the ACET force number. The time average electrothermal force can be written as [75,76]

$$\mathbf{F_E} = -\frac{1}{2} \left[ \left( \frac{\nabla \sigma}{\sigma} - \frac{\nabla \varepsilon}{\varepsilon} \right) \cdot \mathbf{E} \frac{\varepsilon \mathbf{E}}{1 + (\omega \tau)^2} + \frac{1}{2} \left| \mathbf{E} \right|^2 \nabla \varepsilon \right]. \tag{7}$$

where electric field, $\mathbf{E}$ is expressed as $\mathbf{E}(= -\nabla \varphi)$. $\tau = \varepsilon / \sigma$ is the charge relaxation time. Electrothermal forces consist of Coulomb force (first term of the right hand side of the above equation) and dielectric force (second term of the right hand side of the above equation).

To obtain the gradients $\nabla \sigma$ and $\nabla \varepsilon$ we express the functional dependency of $\sigma$ and $\varepsilon$ with temperature and order parameter. On one side, these electrical properties are dependent on temperature distribution, while on the other hand, these are function of order parameter. Considering a linear variation with temperature ($T$),[76] the variation of $\varepsilon$ and $\sigma$ with $T$ and $\phi$ can be written as

$$f(T,\phi) = f_{0,A}\{1+b_A(T-T_0)\}\left(\frac{1+\phi}{2}\right) + f_{0,B}\{1+b_B(T-T_0)\}\left(\frac{1-\phi}{2}\right).$$

Here, parameter $f(T,\phi)$ may be $\sigma$ or $\varepsilon$ and the gradient $b$ may be $\beta$ or $\alpha$. The gradient of permittivity $\alpha$ is order of $-0.001\,\text{K}^{-1}$. Other interfacial properties, such as $\rho, k, C_p$ and $\mu$ are function of only phase field variable $\phi$ and can be written in the following form $f = 0.5 f_A(1+\phi) + 0.5 f_B(1-\phi)$, $f$ may be $\rho, k, C_p$ or $\mu$.

We use the following two boundary conditions for CHE to confirm the zero flux across the rib surfaces and the walls, and to specify wetting characteristics on the walls:[77]

$$\mathbf{n} \cdot \nabla \lambda = 0, \tag{8}$$

$$\mathbf{n} \cdot \nabla \phi = -\tan\left(\frac{\pi}{2} - \theta_d\right)\left|\nabla \phi - (\mathbf{n} \cdot \nabla \phi)\mathbf{n}\right|, \tag{9}$$

Here, $\mathbf{n}$ is the unit normal, outward from the surface and $\theta_d$ is the dynamic contact angle. We impose dynamic contact angle in the form of static contact angle $\theta_s$ via Cox–Voinov equation as:[78,79] $\theta_d^3 = \theta_s^3 + 9Ca\ln(R/l_{slip})$, where R is the macroscopic length scale. $l_{slip}$ is the molecular slip length. The capillary number $Ca$, here, is based on local contact line velocity. Clearly, first boundary condition represents no flux across the solid surfaces. On the other side, second boundary condition controls the profile of the order parameter near the contact line, which, in essence, is the local profile of the interface near the solid surface. Hence, this boundary condition is also known as geometric boundary condition.[77] We have displayed all the boundary conditions in the Fig. 1. A phase difference of 180° is considered between the adjacent electrodes, so that only real part of applied potential needs to be considered. Imposed potential on the electrodes is $\varphi_{rms}^* = \pm 1$, whereas other boundaries are electrically insulating. For thermal field, inlet of the channel and electrodes are set to reference temperature, while at the outlet $\mathbf{n} \cdot \nabla \theta = 0$ is imposed. Other boundaries are insulated. No slip and no penetration boundary conditions are imposed on the walls and ribs to obtain the velocity field. Zero gauge pressure at inlet and outlet is set as the flow is driven by electrical field. At the beginning of the simulations, we have taken $V^* = \theta = (x^*, y^*, t^* = 0) = 0 \,\forall\, x^*, y^*$, as initial conditions. From here onwards, we remove the superscript "*" of the nondimensional parameters for ease of representation of the results.

C. Numerical approach, model benchmarking and mesh independent study

To solve the set of governing equations coupled with the appropriate sets of initial and boundary conditions, we use the platform of COMSOL that is fundamentally based on the Finite Elements Method. The Galerkin least square method is adopted to discretize the convection-diffusion equations. PARDISO solver and the generalized-$\alpha$ scheme are employed for temporal discretization. Tolerance levels of $10^{-6}$ are achieved for all the simulations. We have used extremely fine mesh, where mesh sizes of $\Delta x = \Delta y = 1.25 Cn$ are used for all results. The detailed analysis of the mesh independence study will be discussed later. To confirm an equilibrium and stable interface at the beginning of the simulations the order parameter is initialized by solving the equation $\lambda(\phi) = 0$. The numerical methodologies used in the results are benchmarked with the results of Wang et al.[73] whose details are found in our previous work.[41]

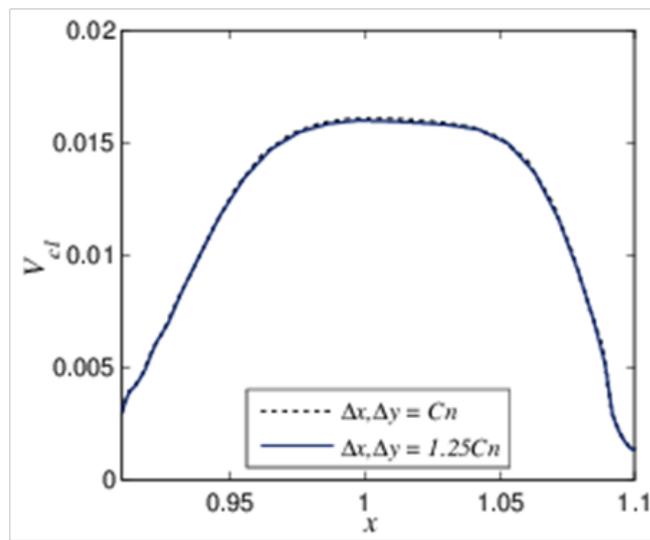

**Fig. 2.** Mesh independent results: The contact line velocity over the first rib is shown for mesh sizes of $\Delta x = \Delta y = Cn$ and $\Delta x = \Delta y = 1.25 Cn$ with $\theta_s = 60°$, $\zeta = 50$, $g = 0.1$, $h = 0.2$, $w = 0.2$ and $x_c = 1$. The figure clearly shows that results are independent with mesh sizes. Therefore, we take mesh size of $\Delta x = \Delta y = 1.25 Cn$ for our numerical result.

To avoid the inaccuracy in the obtained results owing to improper meshing and to achieve a sharp-interface limit, we conducted a mesh independence study. The Cahn number ($Cn = \xi / H$) is defined as the ratio of thin interface thickness to the reference length. Sharp-interface limit will be obtained if $Cn \to 0$.[80] However, it was found that below a critical Cahn number, flow dynamics and order parameter distribution are independent with $Cn$.[80] Hence, Cahn number independent test takes account the mesh independent test. Fig. 2 highlights the mesh independent study wherein we show the variation of contact line velocity over the surface of the first rib for two different mesh sizes of $\Delta x, \Delta y = Cn = 0.01$ and $\Delta x, \Delta y = 1.25 Cn = 0.0125$, the parameters used in the results are shown in the figure caption. It is clear that the deviation of contact line velocity for two different mesh sizes is negligible. Therefore, we take the mesh size of $\Delta x, \Delta y = 1.25 Cn$ for all the results.

## III. Results and discussions

The surface property of the ribs and its geometric parameters have strong effects on alteration of the filling dynamics. Besides, rib arrangement, electrode patterning can have significant effects on the alteration of the flow dynamics. We have thoroughly investigated effects of those parameters. For all the results, the parameters which are constant throughout the analysis are: $Cn = 0.01, Pe_\phi = 0.01, Pe_T = 0.07, \text{Re} = 0.01, Ca = 0.1$, $\zeta = 50, \omega = 0.001, J = 1$ and $B = \eta_T / \eta_0 = -0.01 \text{K}^{-1}$. The values of the dimensionless parameters are evaluated based on the fluid properties of base fluid A, treated as KCl electrolyte solution. Also, we consider property ratios as unity. The length of various dimensions shown in the Fig. 1 are $L = 5, H = 1, d_1 = g = 0.1, d_2 = 0.4, 2s = 0.4$ and $L_1 = 0.5$. Previously reported studies on interfacial motion of two immiscible fluids revealed that flow motion and interfacial dynamics are well described in the form of interface profile, velocity of the fluid-fluid-solid contact line and filling time. Accordingly, we show our investigations in the form of these parameters and interface contours. In addition, to delineate the implications of the numerical results in more details, we represent our observations displaying variation of surface force, showing streamlines and potential lines, in some selected representative cases.

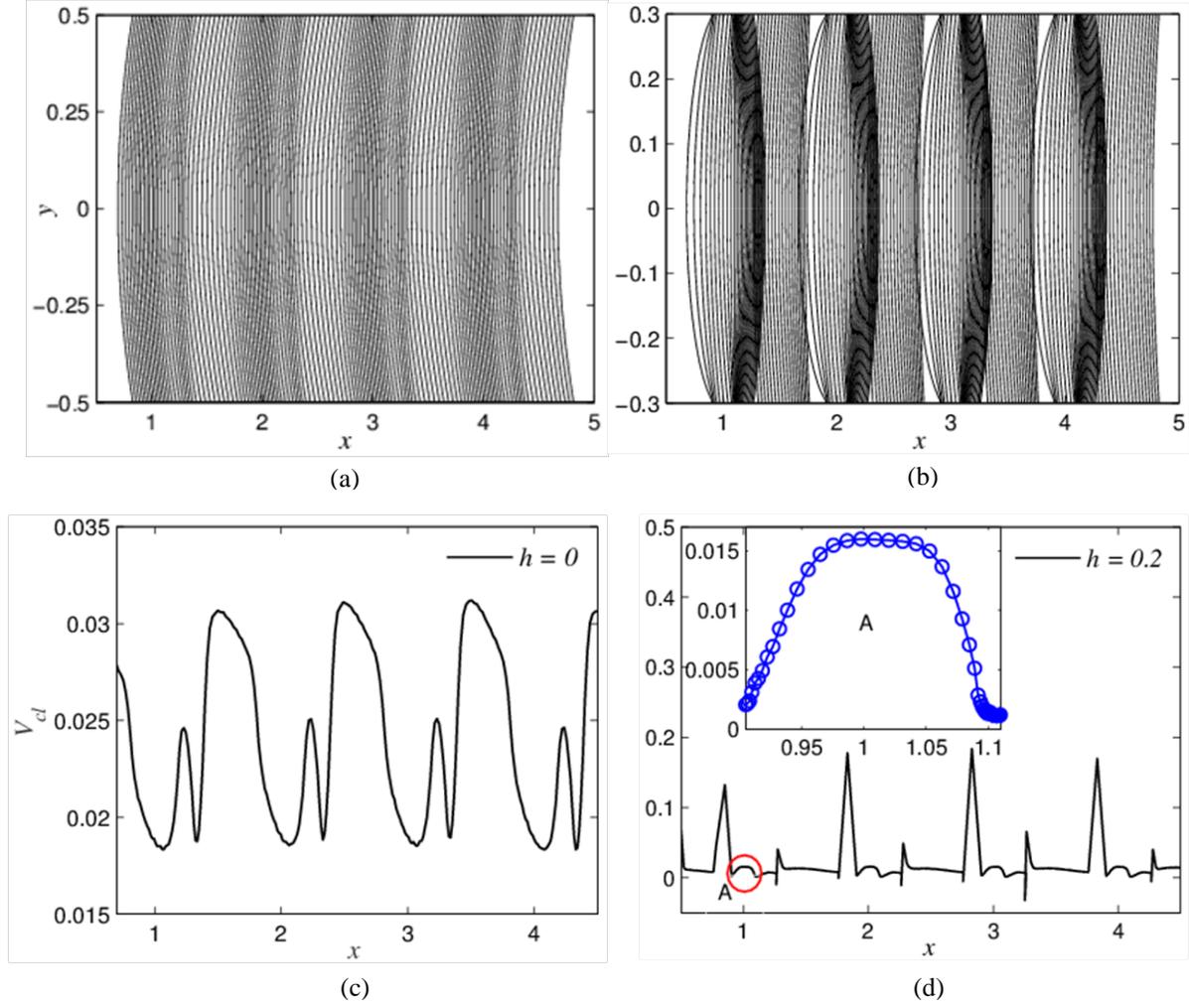

**Fig. 3.** Time evolution of interface contours for (a) channel without rib structure (i.e., $h = 0$), (b) channel with pattern ribs (height $h = 0.2$). Variation of contact line velocity for (c) channel without rib structure (i.e., $h = 0$), (d) channel with pattern ribs (height $h = 0.2$). For all cases, other parameters are $\theta_s = 60°$, $w = 0.2$ and ribs are placed at $x_c = 1, 2, 3$ and 4. Traveling wave like ACET forces results in wavy pattern in the contact line velocity. Interface profiles also show alternate regions of densely and loosely packing. The altered interface pattern with rib structure shows slow traveling of contact line over the rib surface.

We start our observations with a typical comparison between interfacial dynamics of capillary transport for (a) channel without rib structure, (b) channel with rib structure. In order to assess this we show time evolution of interface contours and corresponding contact line velocity for both cases in Fig. (3). Here, we consider rib height as 0.2, the parameters used in the results are given in the figure caption. Point to be noted here that the concept of contact line velocity for case (b) is valid only at positions $x = 0.9 - 1.1, 1.9 - 2.1, 2.9 - 3.1$ and $3.9 - 4.1$, where ribs are placed. At other locations, it is local interface velocity. From the figure it is clear that variation in contact line velocity follows a travelling wave pattern for case (a) with smooth surface. We mentioned earlier that electrothermal forces are generated owing to induced charge in the conducting samples. The gradient of electrical conductivity and permittivity causes induced charge wave which sets motion into the fluid. Electrical signal of travelling potential wave interacts with thermal field and generated ACET forces also become wavy into the bulk. The influence of the traveling wave propagates towards the wall and finally affects the contact line motion. As a consequence, contact line velocity shows pattern of alternate velocity peaks. Interface profiles are also influenced by the traveling wave potential distribution. One can observe that crowded region and commodious region of interface are seen alternately (Fig 3(a)). However, flow pattern and characteristics of contact line dynamics become different when ribs are placed over the channel surface. Contact line moves gently over the rib surface, especially at the trailing edge of the rib (Fig 3(d)). For clear visualization expanded view (marked with red circle) of contact line velocity over the first rib is shown. Contact line gets pinned at the rear end of the rib and almost stops for some time. The details information of flow stoppage can be observed in the plot of interface contours. The densely packed contours show crowding areas just at the end of the rib structure. In the subsequent paragraphs, we reveal the interfacial flow physics in a more involved way considering various characteristics of the rib structure.

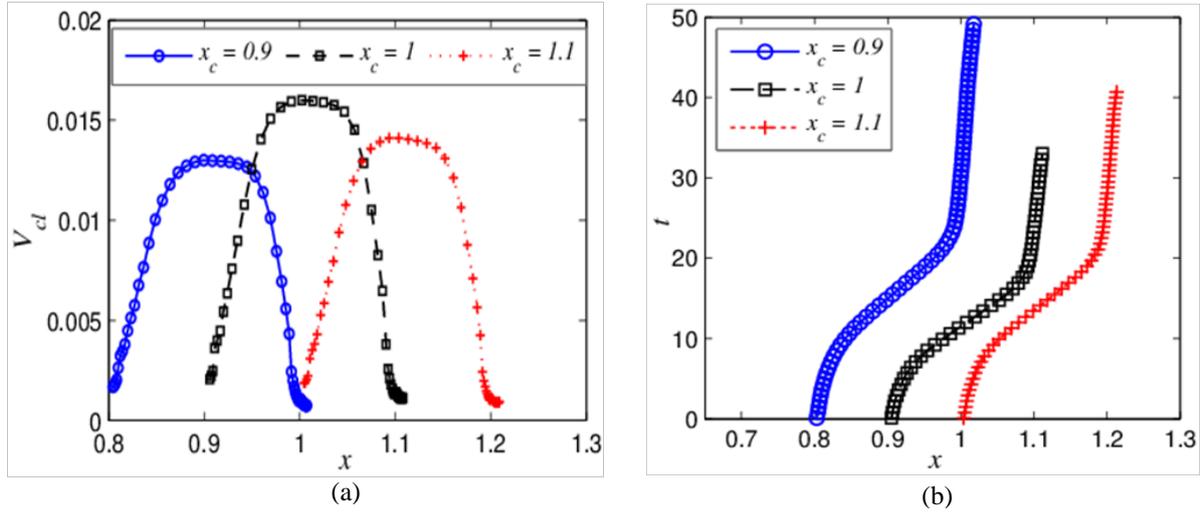

**Fig. 4.** (a) Variation of contact line velocity ($V_{cl}$) on the surface of the first rib with contact line position (x), (b) plot of traversing time ($t$) vs axial position for various locations of the rib ($x_c$); $x_c = 0.9, 1$ and $1.1$. For both the plots other parameters are $\theta_s = 60°$, $h = 0.2$ and $w = 0.2$. The potential distribution which is the most important parameter to cause effective ACET forces gets affected on alteration of the rib locations. We found lower contact line velocity as the ribs are displaced from mid position between the gap of the electrode pair. Accordingly traversing times are also adjusted. For $x_c = 0.9$ we obtain lowest contact line velocity and longest delaying time.

Previously, it was mentioned that due to capillary transport of immiscible fluids interfacial forces become important and interplay with driving ACET forces. The coupling of these forces with driving ACET forces control the flow dynamics of the two phase flow system. We have seen that rib locations change the magnitude of ACET forces, the consequences of which are described in the following. Fig. 4 reveals the underlying contact line dynamics in the form of variation of contact line velocity and time with contact line position for $x_c = 0.9, 1$, and $1.1$ (other parameters used in the study are mentioned in the figure caption). It can be observed that contact line velocity is maximum for $x_c = 1$. Recall from our earlier description of the physical system that a gap of $\Delta x = 2s = 1.2 - 0.8 = 0.4$ is considered between the electrode pairs. Thus, $x_c = 1$ implies that rib is placed at the middle of the gap of the electrode pairs. In the present analysis, we adopt AC electrothermal forces to drive the binary system. ACET forces arise owing to coupling between electrical and thermal fields. The distributions of the potential and temperature deeply affect the magnitude of ACET forces. Symmetrical distributions of these lines generated from equal size electrodes result in vortex motion in the bulk of the fluid. [34,81] The vortex structures of the fluid motion are effective for fluid mixing process. However, to cause unidirectional motion, the vortex pair above the electrodes should be broken. The different methods that can be adopted to ensure this includes, among others, unequal electrodes (asymmetric with respect to mid vertical line of electrodes), employing DC bias on AC electrokinetics, incorporating resistive heater below the electrodes. All of these result in changes in the orientation of the potential lines and thermal distributions, so that a net flow along the channel is observed.[37,54,82,83] In our analysis, we follow the first approach where pairs of asymmetric electrodes are embedded on the top and bottom walls to generate unidirectional flow. When the patterned ribs are placed between the electrode pairs, the potential distribution gets affected. Depending on the position of the ribs, the net driving force is altered and we observe different contact line velocities over the rib surface. When the ribs are shifted from mid position of the gap of electrode pairs along both directions, the contact line velocity decreases. It is noteworthy that contact line velocity for $x_c = 0.9$ (rib after electrode) is higher than the case of $x_c = 1.1$ (rib before electrode). The explanation of intricate flow physics of contact line motion will be explored later showing streamline and potential distributions. Time vs position plot is also an alternative way of presenting the contact line behavior for various positions of the rib. The total time and delaying time exhibit the behavior of contact line velocity with position. For highest velocity (case of $x_c = 1$) the traversing time is lowest. One can see that a large delay time is observed for the location $x_c = 0.9$ for which the contact line velocity was lowest.

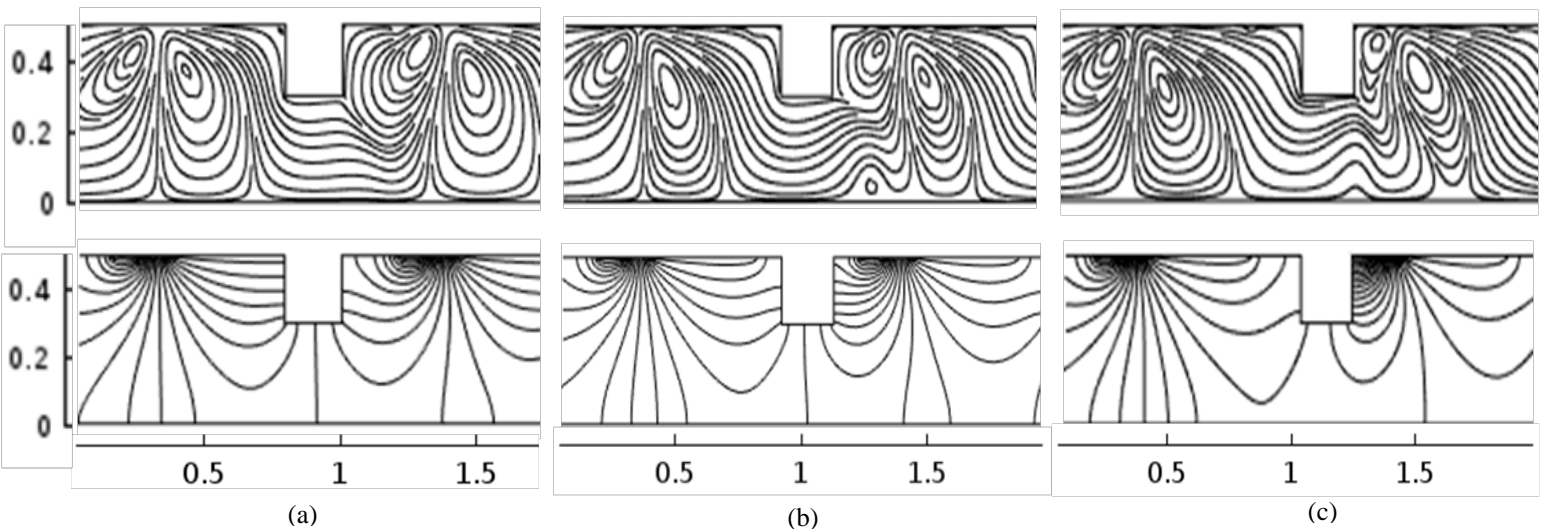

(a)                    (b)                 (c)

**Fig. 5.** (a), (b) and (c) depict potential distributions (on the bottom row) and streamline distributions (on the top row) for $x_c = 0.9, 1$ and 1.1. For all the plots other parameters used in the results are $\theta_s = 60°$, $h = 0.2$ and $w = 0.2$. For $x_c = 0.9$ larger and symmetric vortices hinder the flow passage and result in lower interface velocity. For $x_c = 1.1$ lack of sufficient space behind the narrow width disturbs the potential distribution and lowers the ACET forces.

In Fig. 5, we present the underlying characteristics of potential and flow field on altered ACET forces with various location of the rib. The streamline and potential distributions on the top and bottom row, respectively are displayed for $x_c = 0.9, 1$ and 1.1, at time instance of $t = 80$. One can observe that for $x_c = 0.9$ the rib is located just at the end of the electrode pairs where potential lines are symmetrically oriented and generate two large flow vortices. The interface motion feels a resistance owing to these vortices and slowly progresses on the rib surface. Although the potential is distributed asymmetrically in the flow domain for $x_c = 1.1$, the rib is located adjacent to the narrower electrode. This, in turn, destroys the proper distribution of the potential lines thus lowering the net electrothermal forces. Therefore, ribs located immediately after the electrode pair and before the electrode pair are not suitable for net flow motion. In contrast, when the rib is located at the middle of the gap between the electrode pair it neither modifies the potential lines to form symmetric structure nor disturbs the narrower electrode for complete distribution of potential contours. As a consequence, two vortices (one small and another big) are asymmetrically oriented which do not resist the interface motion and a highest contact line velocity is achieved in this scenario. In summary, we would like to conclude that as per desired conditions, whether it is longer delay time or faster progression of the interface, the arrangement of the ribs can be altered to achieve the outcome.

Figure 6 reveals the influencing characteristics of rib height where we show contact line velocity and time of traveling the first rib with axial location of contact line for various rib height ($h$). The parameters used in the results are mentioned in the figure caption. From the figure, one can see that the contact line velocity gradually decreases as we increase the rib height. As discussed in supplementary material (Fig. S1) that the dissipation is characterized by distortion of the orientation of the interface. The sudden change in direction, position and orientation of the interface results in higher diffusion in a two phase system thereby resulting in alteration of interfacial energy. The surface tension force associated with the surface energy starts opposing the ACET forces over the rib surface. On increasing height, the diffusion becomes higher owing to abruptness in the orientation of the interface and the role of surface tension force in resisting the flow motion increases, thereby slowing down the contact line motion. It may be noted that at lower values of rib height, the deviation of contact line velocity is not high, but it changes rapidly at higher values of rib height. As an example, on changing rib height from $h = 0.1$ to $h = 0.125$ the change in contact line velocity is insignificant. In contrast, it slows down noticeably on changing rib height from $h = 0.175$ to $h = 0.2$.

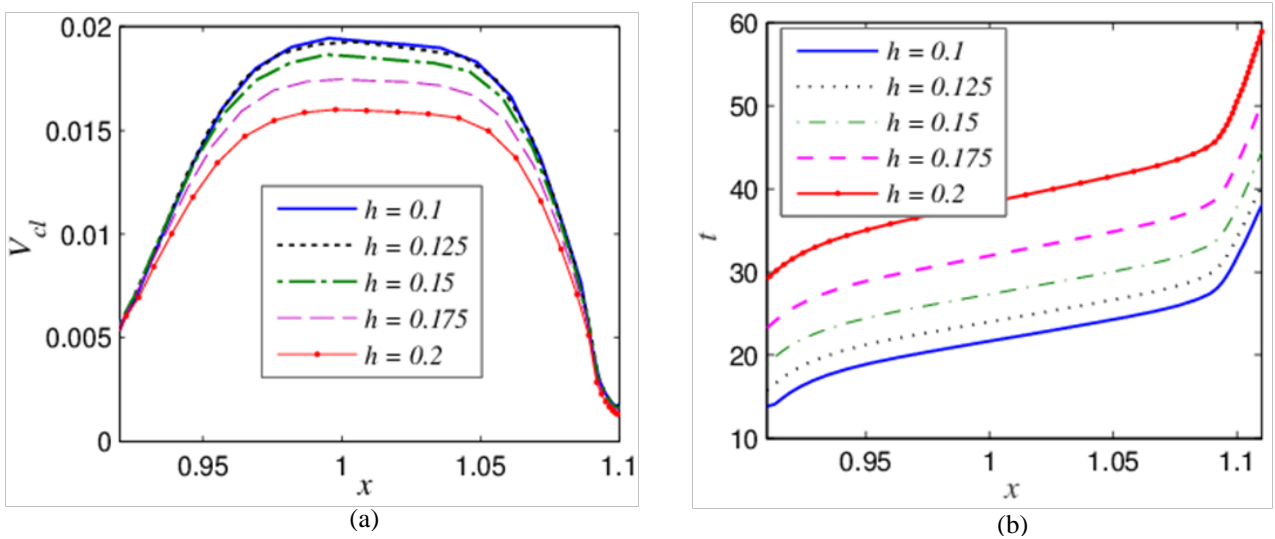

**Fig. 6.** (a) Variation of contact line velocity ($V_{cl}$) on the surface of the first rib with contact line position (x), (b) plot of time of traversing the rib ($t$) vs axial position for $h = 0.1, 0.125, 0.15, 0.175$ and $0.2$. For both the plots other parameters are $\theta_s = 60°, w = 0.2$ and $x_c = 1$. At lower range of rib height, no significant change in contact line velocity is found. However, at higher rib height local change in curvature appreciably changes the diffusion and contact line velocity. Traversing times at the front edge of the rib start at longer time for the higher height of the rib. The lower contact line velocity at higher rib height increases traversing time.

To explore some more important characteristics of the interfacial motion, one can observe the time vs position plot (Fig. 6(b)). The behavior of variation of traveling time is almost identical for all the rib heights. However, the time required for the contact line to reach the front edge is different for different rib heights. This is consistent with our earlier observation where we found the interface to slow down with increasing rib height. From the figure, it can be noticed that delaying time on the trailing edge is higher for higher rib height. At reduced contact line velocity, the edge effect (as discussed previously) is deep which stops the motion for a large time interval on the trailing edge.

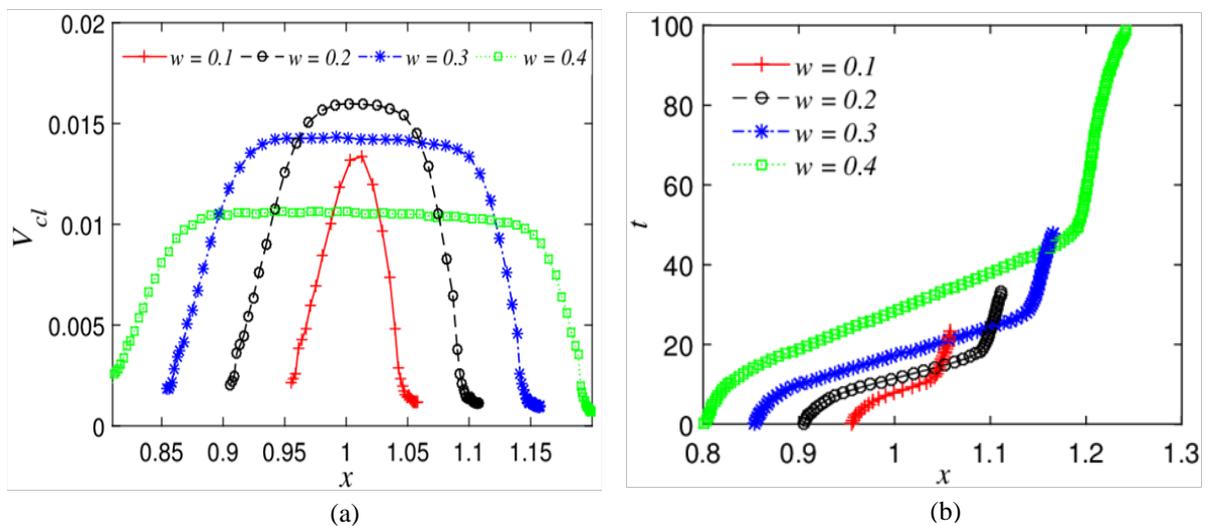

**Fig. 7.** (a) Variation of contact line velocity ($V_{cl}$) on the surface of the first rib with contact line position (x), (b) plot of traveling time ($t$) vs axial position for $w = 0.1, 0.2, 0.3$ and $0.4$. For both the plots, other parameters are $\theta_s = 60°$, $h = 0.2$ and $x_c = 1$. For very small width, the contact line does not have time to show uniform velocity on the rib surface. At larger width, contact line moves with nearly constant velocity for a long time. The surface tension force becomes opposing with net ACET forces over the half width of the rib. Hence, The velocity first increases and then decreases with increasing width. Because of lower velocity at larger width delaying time is long.

In order to analyze the influencing characteristics of width ($w$) of the ribs we display the plots of contact line velocity and time over the first rib with contact line position ($x$) in the Fig. 7 for various rib widths. Important point to be mentioned here is that on alteration of rib width the spacing between two consecutive ribs gets altered as the channel length is fixed. The parameters used in the simulations are given in the figure caption. From Fig. 7(a), one can observe that for very small width, $w = 0.1$ the interface does not have longer time to stay over the rib surface. Therefore, after halting on the front edge the contact line velocity rises to the highest value, then sharply drops to a minimum value. However, with increasing rib width, the contact line travels with constant velocity over the rib surface. The important feature in the context of variation of contact line velocity with rib width is that the peak velocity with which the contact line moves for some time over the surface first increases, then decreases. In the supplementary material, we have mentioned that for wettability of $\theta_s < 90°$ over the half-width the surface tension force aids the interface to move towards the end of the channel, but over another half-width negative surface tension force opposes the interface movement. Here, our investigations involve a surface wettability of $\theta_s = 60°$. Therefore, for wider rib width, after initial adjustment of the various forces, interface has time to traverse with nearly constant velocity. It is therefore clear to note that interplay between surface tension force, ACET force, and viscous force can be controlled with the alteration of rib width to obtain desired contact line velocity. The underlying flow physics of interfacial motion of binary fluids with varying rib width can also be dictated in time vs axial location plot. Surface patterning with ribs with narrower width does not impact the interfacial dynamics much. Since in this case, the retention time is small over the surfaces. We have seen for all widths, at two ends of the rib surface the contact line gets pinned. However, delaying time is small for narrow width patterned. It can be significant increases adopting wider width. Delaying time about $\Delta t \approx 50$ can be achieved with surface patterning rib of width $w = 0.4$.

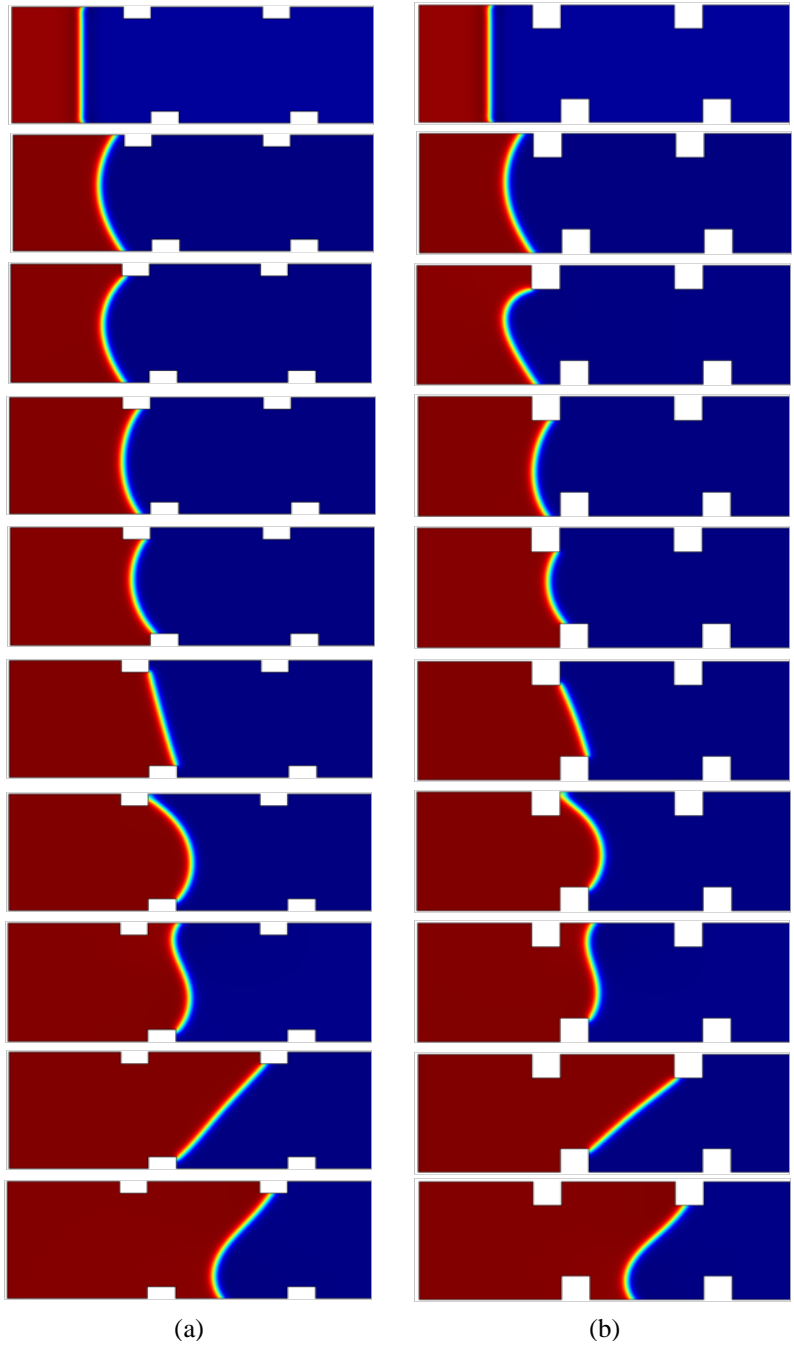

**Fig. 8.** Snapshots show the time evolutions of interface profile with the asymmetric patterning of the ribs wherein ribs on the top wall are places just after the end of the wider electrode (i.e., $x_c = 0.9$) and ribs on the bottom wall are placed just before the narrower electrode (i.e., $x_c = 1.1$). Snapshots on the first column are for (a) $h = 0.1$ whereas for second column (b) $h = 0.2$. For both the cases other parameters are $\theta_s = 60$, $h = 0.2$ and $w = 0.2$. The interface starts with a flat profile. However, pinning-depinning of the interface on the rib surface causes sticking at a different position which, in turn, results in different interface shapes, such as concave, convex, flat with stretching, sickle etc.

In the previous sub-sections, we have investigated the impact of various parameters on contact line motion. Our observations are primarily based on flow control of the binary system and unveiling the intricate dynamics of the interfacial motion. At the beginning, we have seen that pinning and depinning actions of the rib stop the interface at the edges of the rib and speeds up at some other locations. For the symmetrical arrangement of the ribs, it was observed that interface bends toward the advancing fluid A and forms concave shape at the front edge whereas it bends toward receding fluid B at the trailing edge of the rib and makes convex shape. Fig. 8 shows the various interface shapes on progression over the ribs for rib height $h = 0.1$ and 0.2. Here, we place the rib asymmetrically, where the gap between the electrode pairs is filled with two ribs, one at the bottom wall and another on the top wall. Therefore, positions of the ribs are different which make an asymmetric arrangement of the patterned ribs. We show only two pairs of ribs for enlarging visualization. It is clear that initially, the shape of the interface was flat. However, for slight displacement from its initial position, it bends toward the advancing fluid A and forms concave shape. To adjust the wettability condition ($\theta_s = 60°$) imposed on the walls the interface takes this shape. It is to be noted that the interface shape is asymmetric with respect to mid-horizontal line owing to asymmetric flow conditions and boundary conditions. When the one end of the interface touches the edge of the rib it sticks on the edge and severely bends toward the advancing fluid. This phenomenon is highly impacted at higher rib high (see the third instance of the Figs. 8 (a) and (b)). On reaching the trailing edges of the ribs interface shape becomes flat. As time passes the ACET forces push the bulk of the fluid. However, strong pinning action of the trailing edges does not allow ends of the interface to move to the downstream location. As a result, interface bends into the displaced fluid B and form a convex shape in the fluid domain. When the upper end of the interface leaves the edge of the rib located on the top wall it moves faster for a while and crosses a certain distance. Still, the bottom end of the interface is stuck, since the bottom wall rib is at downstream location compared to rib on the top wall. In this scenario, the interface takes S-shape (see the eighth instance of the figure). It is clear that even while the top end of the interface touches next consecutive rib, the bottom end still adheres with the rear end of the bottom wall rib. The interface is highly stretched here. Finally, when the bottom end leaves the rib edge, it moves at faster rate. Still, the other end keeps on pinning motion on the edge of the rib. A sickle shape is thus formed. It can therefore be concluded that asymmetric patterned surface can generate various shape of the advancing front within a short period and small distance.

## IV. Conclusions

Modern microfabrication methodologies have become significantly advanced to fabricate microchannels with internal patterned microstructures that may become useful to control the fluid flow. Keeping this, in mind, we have tried to build fundamental basis and delineate important insights on interfacial dynamics over patterned wetted surfaces of a binary system driven by an AC electrokinetics. We employ phase field method to track the interface and represent the results in the form of interface contours, contact line velocity and traveling time over the first rib on alteration of various tuning parameters. Interfacial behaviour and contact line dynamics of the two phase system are well controlled over the rib surfaces. Delaying time of the interface over the rib surface strongly depends on diffusion. The imposed wettability over the rib surfaces conjunction with various lengths of the rib alter the diffusion rate. As a result, surface tension force becomes important over the rib surfaces. We found that at the edges and nearby locations, pinning motion of the contact line takes place. Interfacial motion almost stops for a long time near the vicinity of the trailing edges. Positions of the ribs have a crucial role on the distribution of the potential lines. As the position of the rib is displaced from mid-position of the gap between electrode pairs, the potential distribution gets altered which generates lesser ACET forces across the contact line and speed of the contact line significantly decreases. Front and rear ends of the rib bend the interface to form concave shape and convex shape. In addition, asymmetric rib patterning changes the topology of the surfaces and results in different shapes of the interface.

We believe that the inferences drawn in the present study on control of flow and shape of the advancing liquid front are expected to be applicable in various microfluidic devices, such as lab-on-a-chip and micro total analysis systems which necessitate a long incubation time or reaction time.

**Electronic supplementary information (ESI)**

Supplementary material related to this article can be found at URL......

23